\documentclass[preprint,12pt]{elsarticle} 
\usepackage{graphicx} 
\usepackage{amssymb} 
\usepackage{amsmath} 

\journal{Nuclear Physics A} 

\begin{document} 

\begin{frontmatter} 

\title{Search for $\eta$-mesic $^{4}\hspace{-0.03cm}\mbox{He}$ in the $dd\rightarrow$ $^{3}\hspace{-0.03cm}\mbox{He} n \pi{}^{0}$ and $dd\rightarrow$ $^{3}\hspace{-0.03cm}\mbox{He} p \pi{}^{-}$ reactions with the WASA-at-COSY facility} 

\author[a]{P.~Adlarson\fnref{fn1}}
\fntext[fn1]{present address: Institut f\"ur Kernphysik, Johannes Gutenberg--Universit\"at Mainz, Johann--Joachim--Becher Weg~45, 55128 Mainz, Germany}
\author[b]{W.~Augustyniak} 
\author[c]{W.~Bardan}     
\author[d]{M.~Bashkanov}  
\author[e]{F.S.~Bergmann}  
\author[f]{M.~Ber{\l}owski}
\author[g]{H.~Bhatt}      
\author[h,i]{A.~Bondar} 
\author[j]{M.~B\"uscher\fnref{fn2}}
\fntext[fn2]{present address:Peter Gr\"unberg Institut, PGI--6 Elektronische 
 Eigenschaften, Forschungszentrum J\"ulich, 52425 J\"ulich, Germany, Institut f\"ur Laser-- und Plasmaphysik, Heinrich--Heine Universit\"at D\"usseldorf, Universit\"atsstr.~1, 40225 D\"usseldorf, Germany}
\author[a]{H.~Cal\'{e}n}   
\author[k]{I.~Ciepa{\l}}  
\author[l,m]{H.~Clement}     
\author[c]{E.~Czerwi{\'n}ski}
\author[e]{K.~Demmich}     
\author[j]{R.~Engels}      
\author[n]{A.~Erven}       
\author[n]{W.~Erven}       
\author[o]{W.~Eyrich}      
\author[j,p]{P.~Fedorets}    
\author[q]{K.~F\"ohl}      
\author[a]{K.~Fransson}    
\author[j]{F.~Goldenbaum}  
\author[j,r]{A.~Goswami}     
\author[j,s]{K.~Grigoryev\fnref{fn3}}
\fntext[fn3]{present address: III.~Physikalisches Institut~B, Physikzentrum, RWTH Aachen, 52056 Aachen, Germany}
\author[a]{C.--O.~Gullstr\"om}
\author[a]{L.~Heijkenskj\"old}
\author[j]{V.~Hejny}      
\author[e]{N.~H\"usken}    
\author[c]{L.~Jarczyk}     
\author[a]{T.~Johansson}   
\author[c]{B.~Kamys}       
\author[t]{N.G.~Kelkar}    
\author[n]{G.~Kemmerling\fnref{fn4}}
\fntext[fn4]{present address: J\"ulich Centre for Neutron Science JCNS, Forschungszentrum J\"ulich, 52425 J\"ulich, Germany}
\author[c]{G.~Khatri\fnref{fn5}}
\fntext[fn5]{present address: Department of Physics, Harvard University, 17~Oxford St., Cambridge, MA~02138, USA}
\author[e]{A.~Khoukaz}     
\author[c]{O.~Khreptak}    
\author[u]{D.A.~Kirillov}  
\author[c]{S.~Kistryn}     
\author[n,4]{H.~Kleines}
\author[v]{B.~K{\l}os}    
\author[f]{W.~Krzemie{\'n}}
\author[k]{P.~Kulessa}    
\author[a,f]{A.~Kup\'{s}\'{c}}
\author[h,i]{A.~Kuzmin}      
\author[w]{K.~Lalwani}     
\author[j]{D.~Lersch}      
\author[j]{B.~Lorentz}     
\author[c]{A.~Magiera}     
\author[j,x]{R.~Maier}       
\author[a]{P.~Marciniewski}
\author[b]{B.~Maria{\'n}ski}
\author[b]{H.--P.~Morsch}  
\author[c]{P.~Moskal}      
\author[j]{H.~Ohm}         
\author[l,m]{E.~Perez del Rio\fnref{fn6}}
\fntext[fn6]{present address: INFN, Laboratori Nazionali di Frascati, Via E. Fermi, 40, 
 00044 Frascati (Roma), Italy}
\author[u]{N.M.~Piskunov} 
\author[j]{D.~Prasuhn}     
\author[a,f]{D.~Pszczel}   
\author[k]{K.~Pysz}       
\author[a,c]{A.~Pyszniak}   
\author[j,x,y]{J.~Ritman}
\author[r]{A.~Roy}        
\author[c]{Z.~Rudy}        
\author[c]{O.~Rundel}      
\author[g]{S.~Sawant}      
\author[j]{S.~Schadmand}  
\author[c]{I.~Sch\"atti--Ozerianska}
\author[j]{T.~Sefzick}    
\author[j]{V.~Serdyuk}     
\author[h,i]{B.~Shwartz}     
\author[e]{K.~Sitterberg} 
\author[l,m,z]{T.~Skorodko}
\author[c]{M.~Skurzok}     
\author[c]{J.~Smyrski}    
\author[p]{V.~Sopov}       
\author[j]{R.~Stassen}    
\author[f]{J.~Stepaniak}   
\author[v]{E.~Stephan}    
\author[j]{G.~Sterzenbach}
\author[j]{H.~Stockhorst}
\author[j,x]{H.~Str\"oher}   
\author[k]{A.~Szczurek}    
\author[b]{A.~Trzci{\'n}ski}
\author[g]{R.~Varma}      
\author[a]{M.~Wolke}       
\author[c]{A.~Wro{\'n}ska} 
\author[n]{P.~W\"ustner}   
\author[xy]{A.~Yamamoto}   
\author[yz]{J.~Zabierowski} 
\author[c]{M.J.~Zieli{\'n}ski}
\author[a]{J.~Z{\l}oma{\'n}czuk}
\author[b]{P.~{\.Z}upra{\'n}ski}
\author[j]{M.~{\.Z}urek}  

\address[a]{Division of Nuclear Physics, Department of Physics and Astronomy, Uppsala University, Box 516, 75120 Uppsala, Sweden}
\address[b]{Department of Nuclear Physics, National Centre for Nuclear Research, ul. Hoza~69, 00-681, Warsaw, Poland}
\address[c]{Institute of Physics, Jagiellonian University, prof. Stanis{\l}awa {\L}ojasiewicza~11, 30-348 Krak\'{o}w, Poland}
\address[d]{School of Physics and Astronomy, University of Edinburgh, James Clerk Maxwell Building, Peter Guthrie Tait Road, Edinburgh EH9 3FD, Great Britain}
\address[e]{Institut f\"ur Kernphysik, Westf\"alische Wilhelms--Universit\"at M\"unster, Wilhelm--Klemm--Str.~9, 48149 M\"unster, Germany}
\address[f]{High Energy Physics Department, National Centre for Nuclear Research, ul. Hoza~69, 00-681, Warsaw, Poland}
\address[g]{Department of Physics, Indian Institute of Technology Bombay, Powai, Mumbai--400076, Maharashtra, India}
\address[h]{Budker Institute of Nuclear Physics of SB RAS, 11~akademika Lavrentieva prospect, Novosibirsk, 630090, Russia}
\address[i]{Novosibirsk State University, 2~Pirogova Str., Novosibirsk, 630090, Russia}
\address[j]{Institut f\"ur Kernphysik, Forschungszentrum J\"ulich, 52425 J\"ulich, Germany}
\address[k]{The Henryk Niewodnicza{\'n}ski Institute of Nuclear Physics, Polish Academy of Sciences,\\52~Radzikowskiego St, 31-342 Krak\'{o}w, Poland}
\address[l]{Physikalisches Institut, Eberhard--Karls--Universit\"at T\"ubingen, Auf der Morgenstelle~14, 72076 T\"ubingen, Germany}
\address[m]{Kepler Center f\"ur Astro-- und Teilchenphysik, Physikalisches Institut der Universit\"at T\"ubingen, Auf der Morgenstelle~14, 72076 T\"ubingen, Germany}
\address[n]{Zentralinstitut f\"ur Engineering, Elektronik und Analytik, Forschungszentrum J\"ulich, 52425 J\"ulich, Germany}
\address[o]{Physikalisches Institut, Friedrich--Alexander--Universit\"at Erlangen--N\"urnberg, Erwin--Rommel-Str.~1, 91058 Erlangen, Germany}
\address[p]{Institute for Theoretical and Experimental Physics, State Scientific Center of the Russian Federation, 25~Bolshaya Cheremushkinskaya, Moscow, 117218, Russia}
\address[q]{II.\ Physikalisches Institut, Justus--Liebig--Universit\"at Gie{\ss}en, Heinrich--Buff--Ring~16, 35392 Giessen, Germany}
\address[r]{Department of Physics, Indian Institute of Technology Indore, Khandwa Road, Indore--452017, Madhya Pradesh, India}
\address[s]{High Energy Physics Division, Petersburg Nuclear Physics Institute, 2~Orlova Rosha, Gatchina, Leningrad district, 188300, Russia}
\address[t]{Departamento de Fisica, Universidad de los Andes, Cra.\~1E, 18A--10, Bogot{\'a}, Colombia}
\address[u]{Veksler and Baldin Laboratory of High Energiy Physics, Joint Institute for Nuclear Physics, 6~Joliot--Curie, Dubna, 141980, Russia}
\address[v]{August Che{\l}kowski Institute of Physics, University of Silesia, Uniwersytecka~4, 40-007, Katowice, Poland}
\address[w]{Department of Physics, Malaviya National Institute of Technology Jaipur, 302017, Rajasthan, India}
\address[x]{JARA--FAME, J\"ulich Aachen Research Alliance, Forschungszentrum J\"ulich, 52425 J\"ulich, and RWTH Aachen, 52056 Aachen, Germany}
\address[y]{Institut f\"ur Experimentalphysik I, Ruhr--Universit\"at Bochum, Universit\"atsstr.~150, 44780 Bochum, Germany}
\address[z]{Department of Physics, Tomsk State University, 36~Lenina Avenue, Tomsk, 634050, Russia}
\address[xy]{High Energy Accelerator Research Organisation KEK, Tsukuba, Ibaraki 305--0801, Japan}
\address[yz]{Department of Astrophysics, National Centre for Nuclear Research, Box 447, 90--950 {\L}\'{o}d\'{z}, Poland}


\begin{abstract} 

The search for $^{4}\hspace{-0.03cm}\mbox{He}$-$\eta$ bound states was performed with the WASA-at-COSY facility via the measurement of the excitation function for the $dd\rightarrow$ $^{3}\hspace{-0.03cm}\mbox{He} n \pi{}^{0}$ and $dd\rightarrow$ $^{3}\hspace{-0.03cm}\mbox{He} p \pi{}^{-}$ processes. The deuteron beam momentum was varied continuously between 2.127~GeV/c and 2.422~GeV/c, corresponding to the excess energy for the $dd\rightarrow$ $^{4}\hspace{-0.03cm}\mbox{He} \eta$ reaction ranging from $Q=$ -70~MeV to $Q=$ 30~MeV. The luminosity was determined based on the $dd\rightarrow$ $^{3}\hspace{-0.03cm}\mbox{He} n$ reaction and the quasi-free proton-proton scattering via $dd\rightarrow$ $p p n_{spectator} n_{spectator}$ reactions. The excitation functions, determined independently for the measured reactions, do not reveal a structure which could be interpreted as a narrow mesic nucleus. Therefore, the upper limits of the total cross sections for the bound state production and decay in $dd\rightarrow(^{4}\hspace{-0.03cm}\mbox{He}$-$\eta)_{bound}\rightarrow$ $^{3}\hspace{-0.03cm}\mbox{He} n \pi{}^{0}$ and $dd\rightarrow(^{4}\hspace{-0.03cm}\mbox{He}$-$\eta)_{bound}\rightarrow$ $^{3}\hspace{-0.03cm}\mbox{He} p \pi{}^{-}$ processes were determined taking into account the isospin relation between the both of the considered channels. 
The results of the analysis depend on the assumptions of the $N^{*}$(1535) momentum distribution in the anticipated mesic-$^{4}\hspace{-0.03cm}\mbox{He}$. Assuming, as in the previous works, that this is identical with the distribution of nucleons bound with 20~MeV in $^{4}\hspace{-0.03cm}\mbox{He}$, we determined that (for the mesic bound state width in the range from 5 MeV to 50 MeV) the upper limits at 90\% confidence level are about 3~nb and about 6~nb for $n\pi^0$ and $p\pi^-$ channels, respectively. 
However, based on the recent theoretical findings of the $N^{*}$(1535) momentum distribution in the $N^{*}$-$^{3}\hspace{-0.03cm}\mbox{He}$ nucleus bound by 3.6 MeV, we find that the WASA-at-COSY detector acceptance decreases and hence the corresponding upper limits are 5~nb and 10~nb for $n\pi^0$ and $p\pi^-$ channels respectively.

\end{abstract}


\end{frontmatter} 

\section{Introduction}
\label{sec:intro}

A possible new kind of exotic nuclear matter called \textit{mesic nucleus} consists of a nucleus bound via the strong interaction with a neutral meson such as the $\eta$, $\eta'$, $K$ or $\omega$ meson.~Some of the most promising candidates for such (unstable) bound states are the \mbox{$\eta$-mesic nuclei}, postulated by Haider and Liu over thirty years ago~\cite{HaiderLiu1}. Current investigations, resulting in a wide range of possible values of the $\eta N$ scattering length, $a_{\eta N}$, determined from hadron- and photon-induced production of the $\eta$ meson, indicate that the attractive $\eta$-nucleon interaction is strong enough to form an $\eta$-nucleus bound system even in light nuclei~\cite{Wilkin1,WycechGreen,Green,Oset2016}.~However, the determination of $\eta$-nucleus scattering length is model dependent and does not permit to claim whether or not a meson binds in nuclei~\cite{AGal}.~Most of the theoretical predictions so far are concerned with nuclei such as carbon or heavier ones, predicting the $\eta$-mesic width in the range of 4-45~MeV~\cite{Garcia2002,HaiderLiu2,HaiderLiu3,Jido,Friedman}. For the $\eta$-mesic $^{4}\hspace{-0.03cm}\mbox{He}$ the predicted width varies in the range of 7-23~MeV~\cite{HaiderLiu2,HaiderLiu3,Sofianos,Rakityansky,Kelk1,Kelk2}.~Therefore in this article we present results of the analysis optimised for the search of the \mbox{$\eta$-mesic} states with the width ranging from 5 to 50~MeV. Moreover, the theories predict $\eta$-nucleus bound states widths which are larger than the binding energies~\cite{Chiang1991,Garcia2002,Oset2016,Rakityansky,Barnea}.~Even though many experimental searches have been carried out until now \cite{Wilkin_Acta,Machner_2015,Kelkar,Kelkar_new,Haider_new,Krusche_Wilkin,Bass_Moskal,Moskal_2016,Moskal_Smyrski}, none of them have brought forth a clear evidence for the existence of such bound states. 

The discovery of mesonic bound states would enable us to broaden the knowledge of the elementary meson-nucleon interaction in the nuclear medium at low energies.~Moreover, it would provide information about the properties of the $\eta$ meson \cite{InoueOset} as well as the $N^{*}(1535)$ resonance~\cite{Jido} inside a nuclear medium.~It could also allow for a better understanding of the $\eta$ and $\eta'$ meson structure, since according to Refs.~\cite{BassTom,BassTomek} the $\eta$ meson binding inside nuclear matter is very sensitive to the singlet component in the quark-gluon wave function of the $\eta$ meson.

It is claimed that a good candidate for the experimental search of possible binding is the \mbox{$^{4}\hspace{-0.03cm}\mbox{He}$-$\eta$} system~\cite{WycechGreen}. Experimental investigations~\cite{Frascaria,Willis} of the interaction between the $^{4}\hspace{-0.03cm}\mbox{He}$ nucleus and the $\eta$ meson lead to observations which suggest the possible existence of the bound state of these two objects~\cite{Wilkin_Acta,Machner_2015}. The production amplitude for the $dd\rightarrow$ $^{4}\hspace{-0.03cm}\mbox{He} \eta$ reaction, extracted from the measured total cross section, rises strongly close to the kinematic threshold. This is a sign of the existence of a pole int the $\eta$-nucleus scattering matrix which can correspond to the bound system.


In June 2008, the WASA-at-COSY collaboration performed an experiment dedicated to search for the $^{4}\hspace{-0.03cm}\mbox{He}$-$\eta$ bound state in the deuteron-deuteron fusion reaction. The experiment was focused on the measurement of the $dd\rightarrow$ $^{3}\hspace{-0.03cm}\mbox{He} p \pi{}^{-}$ reaction in the excess energy range from  \mbox{$Q=$~-51.4~MeV} to $Q=$~22~MeV. The obtained excitation function for this process did not show any resonance like structure which could be interpreted as a signature of \mbox{$\eta$-mesic} $^{4}\hspace{-0.03cm}\mbox{He}$ bound state~\cite{Krzemien_PhD,Adlarson_2013}.~Therefore, an upper limit for the cross-section for the bound state formation and decay in the process $dd\rightarrow$ $(^{4}\hspace{-0.03cm}\mbox{He}$-$\eta)_{bound}$ $\rightarrow$ $^{3}\hspace{-0.03cm}\mbox{He} p \pi{}^{-}$ was determined at the 90\% confidence level and was found to vary from 21 to 27~nb for the assumed width varying from 5~MeV to 45~MeV. 

Here we present results of a subsequent search for the $^{4}\hspace{-0.03cm}\mbox{He}$-$\eta$ state performed with the WASA-at-COSY detector in 2010.~In this new measurement the excess energy range was extended to $Q$ values from -70~MeV to 30~MeV. As compared to the previous experiment~\cite{Adlarson_2013}, the statistics was increased by one order of magnitude and in addition to the $dd\rightarrow$ $^{3}\hspace{-0.03cm}\mbox{He} p \pi{}^{-}$ process, also the $dd\rightarrow$ $^{3}\hspace{-0.03cm}\mbox{He} n \pi^{0}$ reaction was registered~\cite{MSkurzok_PhD, Acta_2015}.~This paper presents the results obtained for the aforementioned processes.

\section{Experiment}

\subsection{Measurement description}

The experiment was performed with high statistics and high acceptance at the 
COSY accelerator using the WASA detection system described in detail 
in Ref.~\cite{Adam}. The WASA detector consists of a Central and Forward part for registering meson decay products and for tagging the recoil particles, 
respectively. The Central Detector consists of the 
drift chamber (straw tubes), plastic scintillators and an electromagnetic calorimeter. 
The momenta of charged particles are determined from the curvature of the 
trajectories in the magnetic field provided by the superconducting solenoid, 
and registered in the straw chamber. The charged particles identification is based on 
the energy deposited by particles in plastic scintillators and in the calorimeter.~The Forward Detector, covering polar angles from 3$^{\circ}$ to 18$^{\circ}$, consists of fourteen planes of plastic 
scintillators and drift tubes which allow for charged particles 
identification and for the track reconstruction, respectively. 

The measurement was carried out with the deuteron COSY beam scattered on an 
internal deuteron pellet target (frozen droplets of deuterium)~\cite{Pellet_target,Pellet_target1}.~During each acceleration cycle, the beam momentum was increased 
continuously from 2.127~GeV/c to 2.422~GeV/c, crossing the kinematic threshold for 
the $dd \rightarrow$ $^{4}\hspace{-0.03cm}\mbox{He} \eta$ reaction at 2.336~GeV/c. 
This range of beam momenta corresponds to the excess energy range $-$70~MeV to 30~MeV. 
The application of this technique allows to reduce systematic 
uncertainties with respect to separate runs at fixed beam energies~\cite{Adlarson_2013, Moskal_Lett, Smyrski, Mersmann}. 

The method used to search for the $\eta$-mesic $^{4}\hspace{-0.03cm}\mbox{He}$ state is based on the measurement of the cross section for the $dd\rightarrow$ $^{3}\hspace{-0.03cm}\mbox{He} n \pi{}^{0}$ and $dd\rightarrow$ $^{3}\hspace{-0.03cm}\mbox{He} p \pi{}^{-}$ processes in the vicinity of the $\eta$ production threshold. If a bound state exists, it should reveal itself as a resonance-like structure in the excitation curve below the $dd\rightarrow$ $^{4}\hspace{-0.03cm}\mbox{He} \eta$ reaction threshold. 
The details of the methods are described in Ref.~\cite{Adlarson_2013, Krzemien_PhD, MSkurzok_PhD}.


\subsection{Identification of $dd\rightarrow$ ($^{4}\hspace{-0.03cm}\mbox{He}$-$\eta)_{bound} \rightarrow$ $^{3}\hspace{-0.03cm}\mbox{He} n \pi{}^{0}$ and $dd\rightarrow$ ($^{4}\hspace{-0.03cm}\mbox{He}$-$\eta)_{bound} \rightarrow$ $^{3}\hspace{-0.03cm}\mbox{He} p \pi{}^{-}$ processes}

\begin{figure}[h!]
\centering
\includegraphics[width=11.5cm,height=6.5cm]{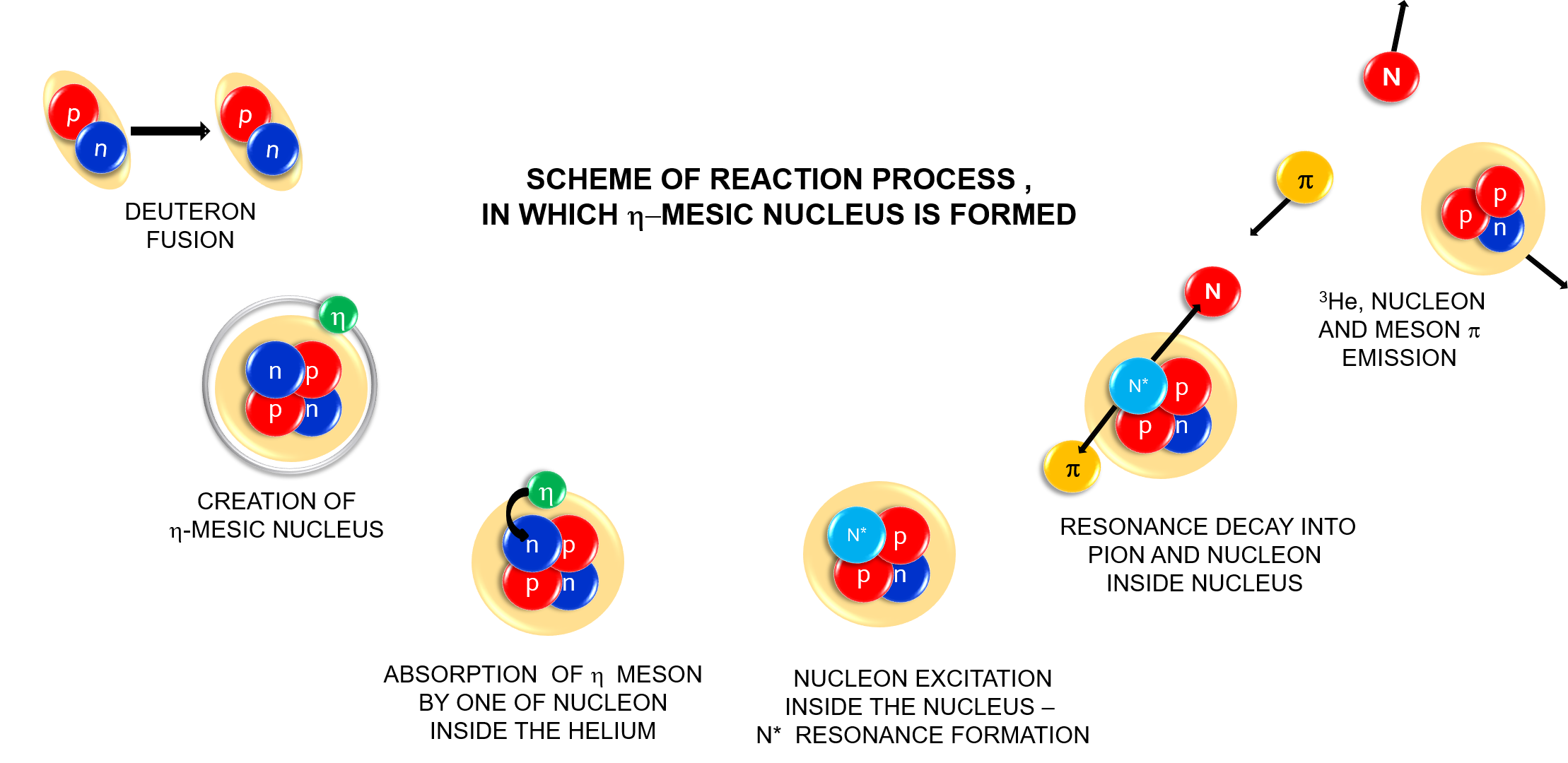}
\caption{Model of the $^{4}\hspace{-0.03cm}\mbox{He}$-$\eta$ bound state production and decay in the \mbox{$dd\rightarrow$ $^{3}\hspace{-0.03cm}\mbox{He} N \pi$} reaction.\label{free_reaction}}
\end{figure}

The selection of the events corresponding to the bound state production in the 
$dd\rightarrow$ $^{3}\hspace{-0.03cm}\mbox{He} N \pi$ reactions was carried out using 
criteria based on the Monte Carlo simulations. The simulations were performed by applying 
the kinematic model of bound state production and decay, schematically presented 
in Fig.~\ref{free_reaction}. 

\vspace{-0.2cm}

As shown in the figure, after the $\eta$-mesic Helium creation in deuteron-deuteron collision, the $\eta$ meson is absorbed on one of the nucleons inside helium and may propagate in the nucleus via consecutive excitations of nucleons to the $N^{*}$(1535) state. The propagation takes place until the resonance decays into the nucleon-pion pair. 
As a first guess, in the simulations, it is assumed that the $N^{*}$ resonance in the 
center of mass frame moves with a momentum distribution similar to that of nucleons inside $^{4}\hspace{-0.03cm}\mbox{He}$~\cite{Nogga,Nogga2}. This assumption was used in the previous work~\cite{Adlarson_2013}. In addition, in this analysis the simulations were also performed assuming the momentum distribution of the $N^{*}$ in the  $N^{*}$-$^{3}\hspace{-0.03cm}\mbox{He}$ bound state according to the very recent theoretical appraisals from references~\cite{Kelkar_2015_new, Kelkar_2016_new}.

~The $^{3}\hspace{-0.03cm}\mbox{He}$ nucleus, consisting of three other nucleons, plays then a role of a spectator. The simulations were carried out under the assumption that the bound state has a Breit-Wigner resonance structure with fixed binding energy $B_{s}$ and a width $\Gamma$. The deuteron beam momentum was generated with a uniform probability density distribution in the range of $p_{beam}\in$(2.127,2.422)GeV/c which corresponds to the experimental beam ramping.

\begin{figure}[h]
\centering
\includegraphics[width=9.0cm,height=6.0cm]{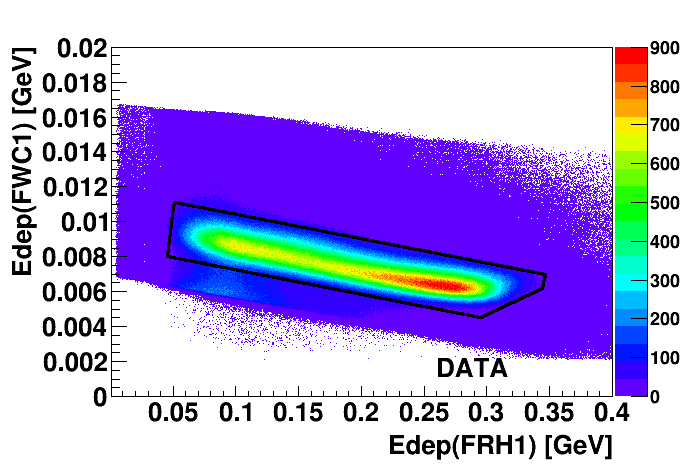} 
\caption{Spectrum of energy deposited in the first layer of Forward Window Counter (FWC1) and the first layer of Forward Range Hodoscope (FRH1) for experimental data. The selected area for $^{3}\hspace{-0.03cm}\mbox{He}$ is marked with black line. The empty area below comes from the preselection conditions.\label{Hel_ident}}
\end{figure}

Analyses of the $dd\rightarrow$ $^{3}\hspace{-0.03cm}\mbox{He} n \pi{}^{0}$ and $dd\rightarrow$ $^{3}\hspace{-0.03cm}\mbox{He} p \pi{}^{-}$ reactions were carried
out independently. The Helium ions and nucleon-pion pairs were registered in the Forward and Central Detector, respectively. The $^{3}\hspace{-0.03cm}\mbox{He}$ identification was carried out with $\Delta$E-E method based on energy losses in scintillator layers of the Forward Detector (see e.g. Fig.~\ref{Hel_ident}). 

\vspace{-0.3cm}

\begin{figure}[h!]
\centering
\includegraphics[width=6.5cm,height=4.5cm]{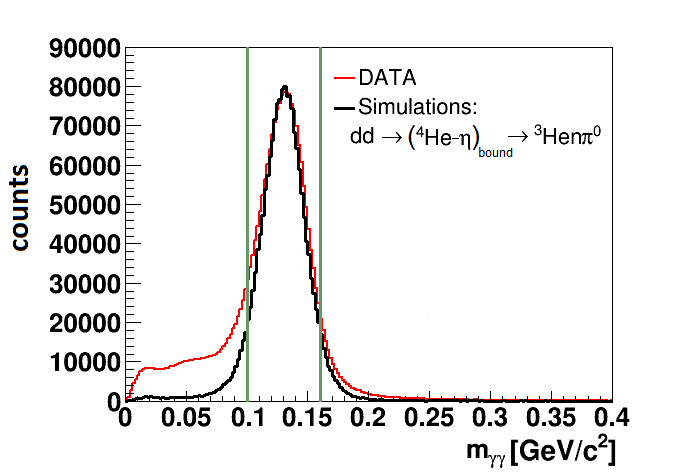}
\includegraphics[width=6.5cm,height=4.5cm]{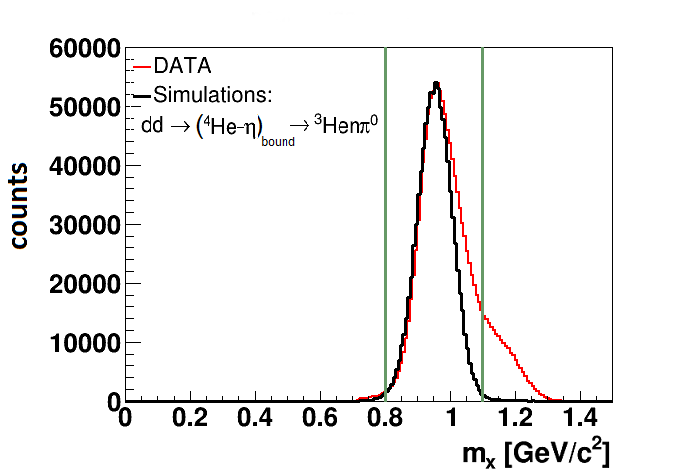}
\caption{$\pi^{0}$ identification based on invariant mass spectrum (left panel) and neutron identification via the missing mass technique (right panel).~Green vertical lines indicate the boundary of the applied selection criteria.\label{pion_ident}}  
\end{figure}

Proton and $\pi^{-}$ identification is based on the energy loss in the Plastic Scintillator combined with the energy deposited in the Electromagnetic Calorimeter and is described in details in Ref.~\cite{Krzemien_PhD, Adlarson_2013}. The neutral pions $\pi^{0}$ four-vectors were reconstructed by combining the four-vectors of gamma quanta pairs registered in the Calorimeter and selected under the condition imposed on their invariant mass, while the missing mass technique allowed to identify neutrons~\cite{MSkurzok_PhD}. The invariant and missing mass spectra are shown in Fig.~\ref{pion_ident}.    

\vspace{-0.2cm}

\begin{figure}[h!]
\centering
\includegraphics[width=6.5cm,height=4.5cm]{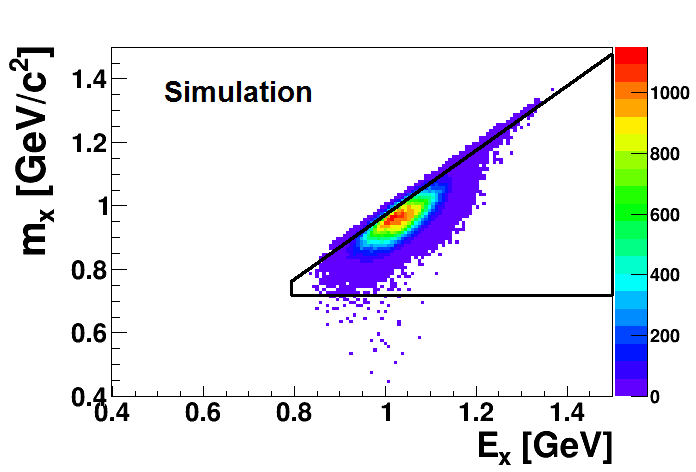}
\includegraphics[width=6.5cm,height=4.5cm]{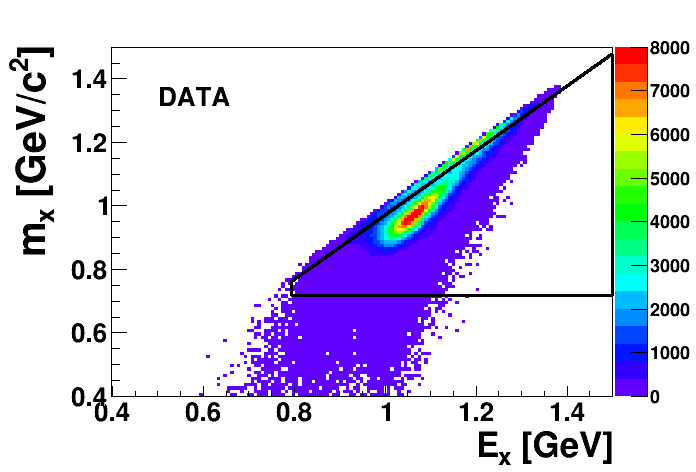}
\caption{Missing mass $m_{x}$ vs. missing energy $E_{x}$ for Monte Carlo simulation of the $dd\rightarrow$ ($^{4}\hspace{-0.03cm}\mbox{He}$-$\eta)_{bound} \rightarrow$ $^{3}\hspace{-0.03cm}\mbox{He} n \pi{}^{0}$ (left panel) and experimental data (right panel). The applied graphical condition is marked as black solid curve. \label{mx_Ex_main}}
\end{figure}

Additional criteria applied in $m_{x}(E_{x})$ spectrum, shown in Fig.~\ref{mx_Ex_main}, allowed to reduce the background coming from multipion processes - mainly from the $dd \rightarrow$ $^{3}\hspace{-0.03cm}\mbox{He} n \pi^{0} \pi^{0}$ reaction. 

\begin{figure}[h!]
\centering
\includegraphics[width=6.5cm,height=4.4cm]{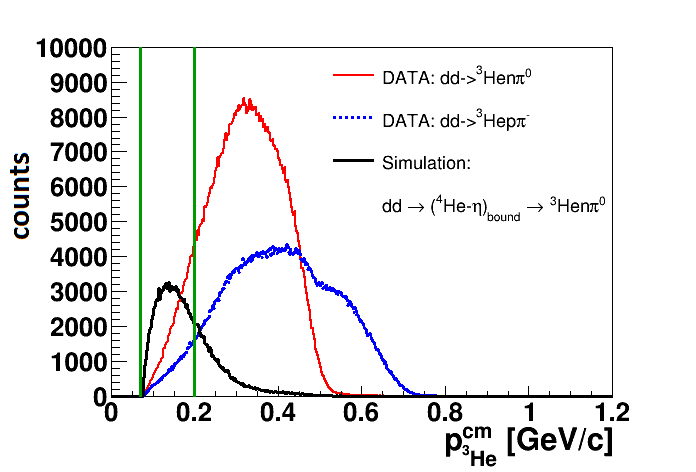} 
\includegraphics[width=6.5cm,height=4.4cm]{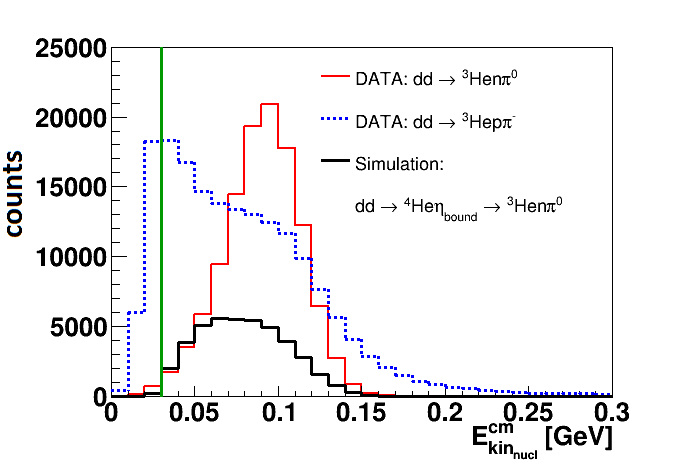}\\

\includegraphics[width=6.5cm,height=4.4cm]{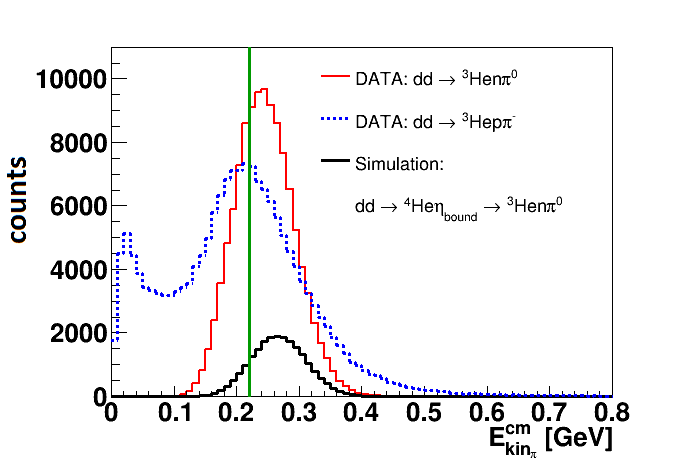}
\includegraphics[width=6.5cm,height=4.4cm]{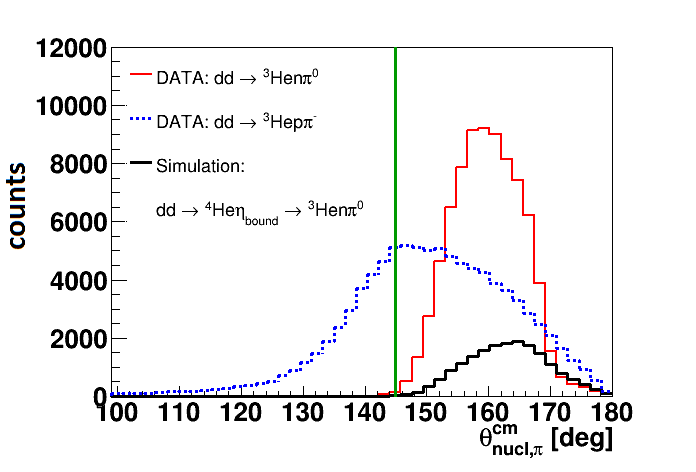}

\caption{Spectrum of $^{3}\hspace{-0.03cm}\mbox{He}$ center-of-mass momentum $p^{cm}_{^{3}\hspace{-0.05cm}He}$ (left upper panel), center-of-mass kinetic energy of nucleon $E^{cm}_{kin_{nucl}}$ (right upper panel), center-of-mass kinetic energy of pion $E^{cm}_{kin_{\pi}}$ (left lower panel) and nucleon-pion opening angle in the center-of-mass $\theta^{cm}_{nucl,\pi}$ (right lower panel). Data are shown in red (thin solid line) and blue (dotted) line for $dd\rightarrow$ $^{3}\hspace{-0.03cm}\mbox{He} n \pi{}^{0}$ and $dd\rightarrow$ $^{3}\hspace{-0.03cm}\mbox{He} p \pi{}^{-}$ reaction, respectively. Monte Carlo simulations of signal normalized arbitrarily are shown in black (thick solid line), while the green vertical lines indicate the boundary of the applied selection criteria.\label{kinem_cuts}}
\end{figure}

Events corresponding to the production of the $\eta$-mesic Helium were selected for both 
the considered reactions based on the $^{3}\hspace{-0.03cm}\mbox{He}$ momentum in the center of mass. The signal rich region corresponds to the center of mass momenta of the $^{3}\hspace{-0.03cm}\mbox{He}$ in the range of \mbox{$p^{cm}_{^{3}\hspace{-0.05cm}He}\in(0.07,0.2)$~GeV/c}. The selection was improved by additional criteria using the nucleon and pion kinetic energies as well as the nucleon-$\pi$ opening angle in the center of mass system. The spectra with marked boundaries are presented in Fig.~\ref{kinem_cuts}.

\begin{figure}[h]
\centering
\includegraphics[width=8.2cm,height=5.2cm]{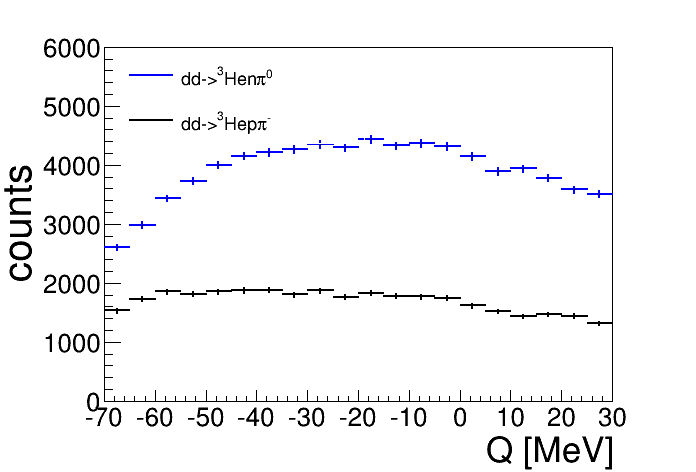}
\vspace{-0.2cm}
\caption{Raw excitation function for the $dd\rightarrow$ $^{3}\hspace{-0.03cm}\mbox{He} n \pi{}^{0}$ reaction (blue) and  the \mbox{$dd\rightarrow$ $^{3}\hspace{-0.03cm}\mbox{He} p \pi{}^{-}$} reaction (black). The shown spectra are not corrected for efficiency and luminosity. Horizontal bars indicate the size of the excess energy interval equal to 5~MeV.~\label{exc_fcn_allcuts}}

\end{figure}

The yields of the selected events for both processes are shown in Fig.~\ref{exc_fcn_allcuts} as a function of the excess energy.

\subsection{Efficiency}

\begin{figure}[h!]
\includegraphics[width=7.0cm,height=4.3cm]{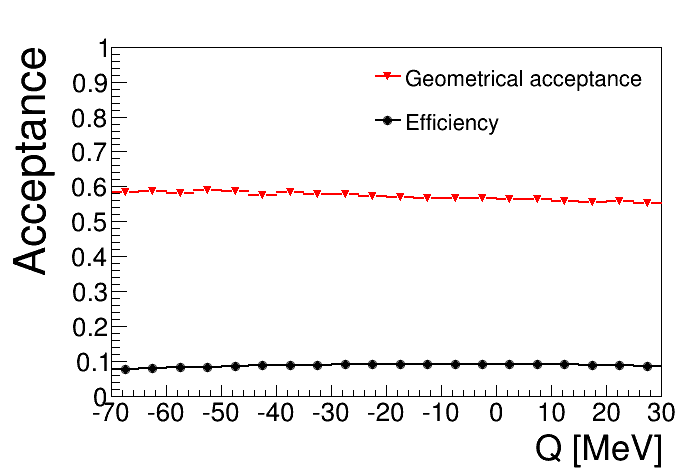} 
\includegraphics[width=7.0cm,height=4.3cm]{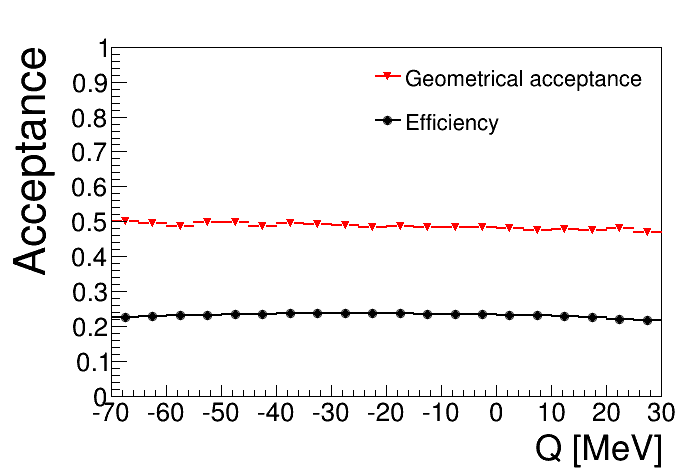} 

\caption{The acceptance and efficiency for $dd\rightarrow$ $(^{4}\hspace{-0.03cm}\mbox{He}$-$\eta)_{bound} \rightarrow$ $^{3}\hspace{-0.03cm}\mbox{He} n \pi^{0}$ (left panel) and $dd\rightarrow$ $(^{4}\hspace{-0.03cm}\mbox{He}$-$\eta)_{bound} \rightarrow$ $^{3}\hspace{-0.03cm}\mbox{He} p \pi^{-}$ (right panel) reactions as a function of excess energy $Q$. The geometrical acceptance of the WASA detector for both channels is shown with red triangles while the full efficiency including detection and reconstruction efficiency for the region rich in signal is shown with black circles. Horizontal bars indicate the size of the excess energy interval and statistical errors (hardly visible) are shown by vertical bars.~\label{acc_eff}}
\end{figure}

\indent The overall detection and reconstruction efficiencies were determined based on the Monte Carlo simulation for the $dd\rightarrow$ ($^{4}\hspace{-0.03cm}\mbox{He}$-$\eta)_{bound} \rightarrow$ $^{3}\hspace{-0.03cm}\mbox{He} n \pi{}^{0}$ and $dd\rightarrow$ ($^{4}\hspace{-0.03cm}\mbox{He}$-$\eta)_{bound} \rightarrow$ $^{3}\hspace{-0.03cm}\mbox{He} p \pi{}^{-}$ processes taking into account response of detection system and selection criteria applied in the data analysis and the Fermi momentum distribution of nucleons in the $^{4}\hspace{-0.03cm}\mbox{He}$ nucleus according to the model in ~\cite{Nogga}. The efficiency was calculated for each of the excess energy intervals as a ratio of the number of reconstructed to the generated events. For the generated events a detector response was simulated and the analysis was conducted taking into account the same selection criteria as for the experimental data. Fig.~\ref{acc_eff} shows the efficiency for the region rich in signal as well as the WASA detector acceptance for both the reactions. 


More detailed investigations showed that the efficiency dependences on the bound state width $\Gamma$ and the binding energy $B_{s}$ is negligible~\cite{MSkurzok_PhD} if the Fermi momentum distribution is simulated according to the model from Ref.~\cite{Nogga}.

\subsection{Luminosity}

\begin{figure}[h!]
\centering
\includegraphics[width=8.5cm,height=5.2cm]{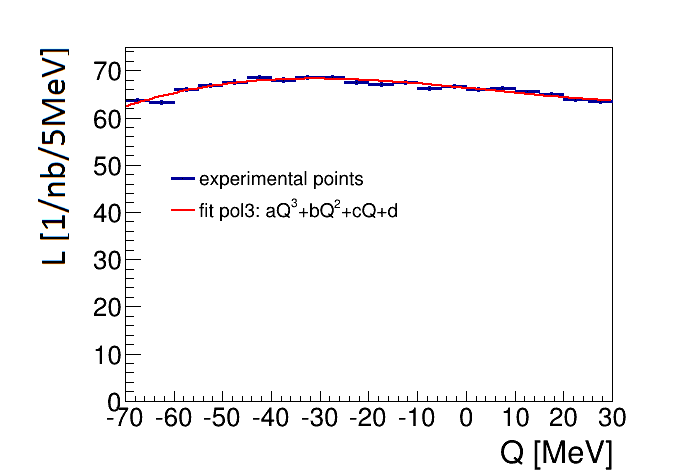}
\vspace{-0.2cm}
\caption{Integrated luminosity calculated for the experimental data for the quasi-free $dd\rightarrow p p n_{spectator} n_{spectator}$ reaction (blue points). The superimposed red solid line indicates a result of the fit of the third degree polynomial function. Horizontal bars indicate the size of the excess energy interval and statistical errors (hardly visible) are shown by vertical bars.~\label{lum_norm_fit}}
\end{figure}

During the beam ramp cycle, the luminosity changes due to beam losses, as well as due to the changes in the beam-target overlap and adiabatic beam size shrinking~\cite{Lorentz}.~Therefore, both the total integrated luminosity (i.e. integrated luminosity summed up over cases of different excess energy values) and the dependence of integrated luminosity on the excess energy has to be determined. The total integrated luminosity was calculated based on two reactions: $dd\rightarrow$ $^{3}\hspace{-0.03cm}\mbox{He} n$ and $dd\rightarrow p p n_{spectator} n_{spectator}$. The integrated luminosity dependence on the excess energy, used for normalization of the excitation functions, was determined based on quasi free $dd\rightarrow p p n_{spectator} n_{spectator}$ reaction and is presented in Fig.~\ref{lum_norm_fit}.


Total integrated luminosities obtained for the two aforementioned processes are consistent within systematics and normalization errors and are equal to $L^{tot}_{dd\rightarrow ^{3}\hspace{-0.03cm}He n}=(1102\pm2_{stat}\pm28_{syst}\pm107_{norm})$~nb$^{-1}$ and $L^{tot}_{dd\rightarrow p p n_{spectator} n_{spectator}}$ =$(1326\pm2_{stat}\pm108_{syst}\pm64_{norm})$~nb$^{-1}$, respectively. The detailed description of the luminosity determination can be found in Ref.~\cite{MSkurzok_PhD, Acta_2015}. It is worth emphasizing that both efficiency (Fig.~\ref{acc_eff}) and luminosity (Fig.~\ref{lum_norm_fit}) are smooth functions of the excess energy which allows to avoid any artefact structures in the determined cross section spectrum.


\section{Upper limits of the total cross section}

\begin{figure}[h!]
\centering
\includegraphics[width=6.5cm,height=4.5cm]{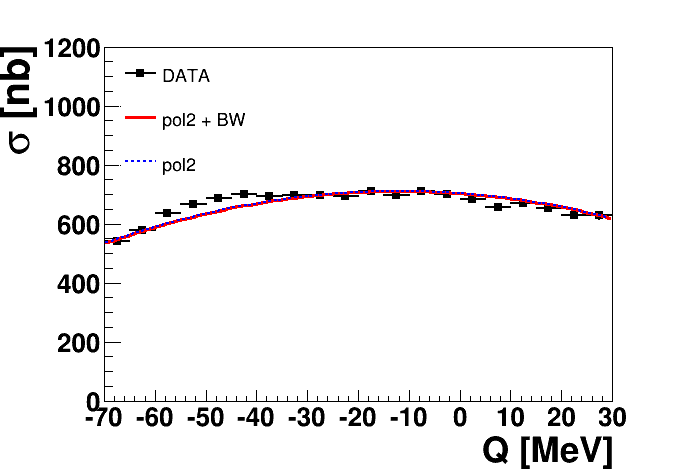}
\includegraphics[width=6.5cm,height=4.5cm]{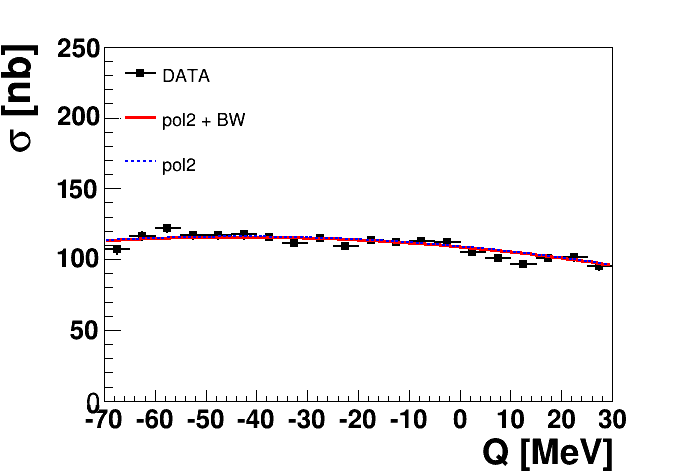}
\vspace{-0.2cm}
\caption{Excitation function for the $dd\rightarrow$ $^{3}\hspace{-0.03cm}\mbox{He} n \pi{}^{0}$ reaction (left panel) and  the \mbox{$dd\rightarrow$ $^{3}\hspace{-0.03cm}\mbox{He} p \pi{}^{-}$} reaction (right panel). The red solid line represents a fit with a second order polynomial combined with a Breit-Wigner function with fixed binding energy and width equal to 30 and 40~MeV, respectively. The blue dotted line shows the second order polynomial describing the background. Horizontal bars indicate the size of the excess energy interval and statistical errors (hardly visible) are shown by vertical bars.~\label{ex_fcn_fit}}
\end{figure}

The excitation functions of the total cross section for both investigated processes (Fig.~\ref{ex_fcn_fit}) were determined for the region rich in signal by dividing the number of events in each excess energy interval (Fig.~\ref{exc_fcn_allcuts}) by the corresponding integrated luminosity $L(Q)$ (Fig.~\ref{lum_norm_fit}) and correcting for the efficiency (Fig.~\ref{acc_eff}).~The obtained excitation curves do not show any structure for energies below the $\eta$ production threshold which could be the signature of the narrow $^{4}\hspace{-0.03cm}\mbox{He}$-$\eta$ bound state existence.~Therefore, an upper limit for the cross-section for formation of the $^{4}\hspace{-0.03cm}\mbox{He}$-$\eta$ bound state and its decay into the $^{3}\hspace{-0.03cm}\mbox{He} n \pi{}^{0}$ and $^{3}\hspace{-0.03cm}\mbox{He} p \pi{}^{-}$ channels were calculated. The excitation functions for both processes were fitted simultaneously with a sum of a second order polynomial and a Breit-Wigner function describing the background and the signal from the bound state, respectively. Thereby the isospin relation between $n \pi{}^{0}$ and $p \pi{}^{-}$ pairs emerging from the $N^{*}$ decay has been taken into account, which state that the probability of $p \pi{}^{-}$ pair production is two times higher than in case of $n \pi{}^{0}$ production. The fit was conducted with fixed binding energy $B_{s}$ in the range from 0 to 40 MeV and bound state width $\Gamma$ from 5 to 50~MeV, while the polynomial coefficients and the normalization of the Breit-Wigner amplitude were treated as free parameters. As an example, the excitation functions with the fit results for binding energy 30~MeV and width 40~MeV are presented in Fig.~\ref{ex_fcn_fit}.

There are 4$\sigma$ indications of structures above background in the case of the $dd\rightarrow$ $^{3}\hspace{-0.03cm}\mbox{He} n \pi{}^{0}$ channel, however their assignment to the mesic state is excluded by the comparison with the $dd\rightarrow$ $^{3}\hspace{-0.03cm}\mbox{He} p \pi^{-}$. The simultaneous fit to both channels gives a Breit-Wigner contribution consistent with zero within 2$\sigma$. Therefore, the upper limit of the total cross section was calculated at the confidence level 90\% based on standard deviation $\sigma_{A}$ of the incoherent square of the Breit-Wigner amplitude obtained from the fit ($\sigma^{upp}_{CL=90\%}=k \cdot \sigma_{A}$ with $k=1.64$ as given in PDG~\cite{pdg}). The values of the obtained upper limits are shown in Table~\ref{tab_upp_lim}.

\begin{table}[h!]
\begin{footnotesize}
\begin{center}
\begin{tabular}{|c|c|c||c|c|c|}
\hline
$B_{s}$~[MeV] & $\Gamma$~[MeV] & $\sigma^{upp}_{90\%}$ [nb] &$B_{s}$~[MeV] & $\Gamma$~[MeV] & $\sigma^{upp}_{90\%}$ [nb]\\
\hline
\hline

10 &5 &3.8 &30 &5 &3.8 \\
10 &10 &2.6 &30 &10 &2.5\\
10 &20 &2.6 &30 &20 &2.4 \\
10 &30 &3.1 &30 &30 &2.6\\
10 &40 &3.8 &30 &40 &3.1\\
10 &50 &4.8 &30 &50 &3.7\\
20 &5 &3.9 &40 &5 &3.9\\
20 &10 &2.6 &40 &10 &2.6\\
20 &20 &2.6 &40 &20 &2.4\\
20 &30 &3.0 &40 &30 &2.7\\
20 &40 &3.7 &40 &40 &3.1\\
20 &50 &4.7 &40 &50 &3.7\\

\hline 
\end{tabular}
\end{center}
\vspace{-0.5cm}
\begin{center}
\caption{The upper limit of the total cross-section for the $dd\rightarrow$ ($^{4}\hspace{-0.03cm}\mbox{He}$-$\eta)_{bound} \rightarrow$ $^{3}\hspace{-0.03cm}\mbox{He} n \pi{}^{0}$ process determined at CL=90\% for different values of binding energy $B_{s}$ and width $\Gamma$. The upper limit of the total cross-section for the $dd\rightarrow$ ($^{4}\hspace{-0.03cm}\mbox{He}$-$\eta)_{bound} \rightarrow$ $^{3}\hspace{-0.03cm}\mbox{He} p \pi{}^{-}$ process according to isospin relation is two times larger.\label{tab_upp_lim}}
\end{center}
\end{footnotesize}
\end{table}

\begin{figure}[h!]
\centering
\includegraphics[width=6.5cm,height=4.5cm]{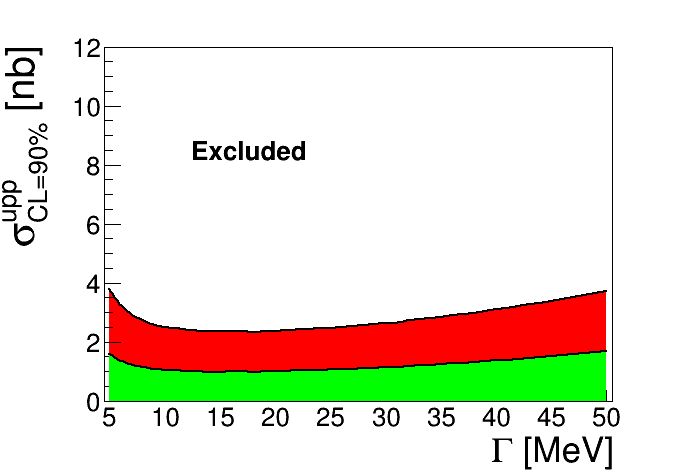}
\includegraphics[width=6.5cm,height=4.5cm]{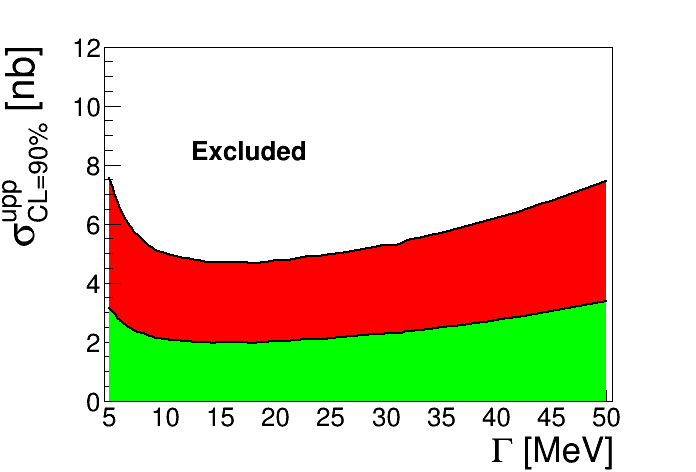}
\vspace{-0.2cm}
\caption{Upper limit of the total cross-section for $dd\rightarrow(^{4}\hspace{-0.03cm}\mbox{He}$-$\eta)_{bound}\rightarrow$ $^{3}\hspace{-0.03cm}\mbox{He} n \pi{}^{0}$ (upper panel) and $dd\rightarrow(^{4}\hspace{-0.03cm}\mbox{He}$-$\eta)_{bound}\rightarrow$ $^{3}\hspace{-0.03cm}\mbox{He} p \pi{}^{-}$ (lower panel) reaction as a function of the width of the bound state. The binding energy was fixed to 30~MeV. The upper limit was determined via the simultaneous fit for both channels. The green area denotes the systematic uncertainties.~\label{Result_sigma_upp_both}}  
\end{figure}

It is worth emphasizing that the upper limit depends mainly on the bound state width and just slightly changes with the binding energy (for the  analysis  done under assumption of the Fermi momentum distribution as given in reference~\cite{Nogga}).~The obtained upper limits as a function of the bound state width are presented for each of the studied reactions in Fig.~\ref{Result_sigma_upp_both}. 

They vary from 2.5 to 3.5 nb for the first process and from 5 to 7 nb for the second process for the width ranging from 5 to 50 MeV. The values of the achieved upper limits are predominantly due to the $^{3}\hspace{-0.03cm}\mbox{He} p \pi{}^{-}$ channel since the background for this channel is about six smaller than the background due to the $dd\rightarrow$ $^{3}\hspace{-0.03cm}\mbox{He} n \pi{}^{0}$  process (Fig.~\ref{ex_fcn_fit}). The green area denotes the systematics errors described in the next section.

\section{Systematics}

Systematic studies were carried out analogically to previous analysis described in~Ref.~\cite{Adlarson_2013}. It was investigated, how the variation of the selection criteria and application of different theoretical models and assumptions influences the obtained result.

The variation of the selection conditions by $\pm$10\% results in the systematic error of about 6\% in case of $dd\rightarrow(^{4}\hspace{-0.03cm}\mbox{He}$-$\eta)_{bound}\rightarrow$ $^{3}\hspace{-0.03cm}\mbox{He} n \pi{}^{0}$ and $dd\rightarrow(^{4}\hspace{-0.03cm}\mbox{He}$-$\eta)_{bound}\rightarrow$ $^{3}\hspace{-0.03cm}\mbox{He} p \pi{}^{-}$ reactions.

A significant source of the systematic error of the upper limit is related to the \mbox{luminosity} determination based on the quasi-free $pp$ reaction. The systematic and normalization \mbox{luminosity} errors are equal to about 8\% and 5\%, respectively. The details of the \mbox{luminosity} systematics analysis can be found in Ref.~\cite{MSkurzok_PhD}. 

The fitting assumptions applied in the analysis provide additional uncertainty.~The error from the fit of a quadratic or linear function to the background is estimated as $\frac{\sigma_{quad}-\sigma_{lin}}{2}$. It changes from about 3\% ($\Gamma$=5~MeV) to 18\% ($\Gamma$=50~MeV) for both of considered process. 

\begin{figure}[h!]
\centering
\includegraphics[width=8.0cm,height=6.0cm]{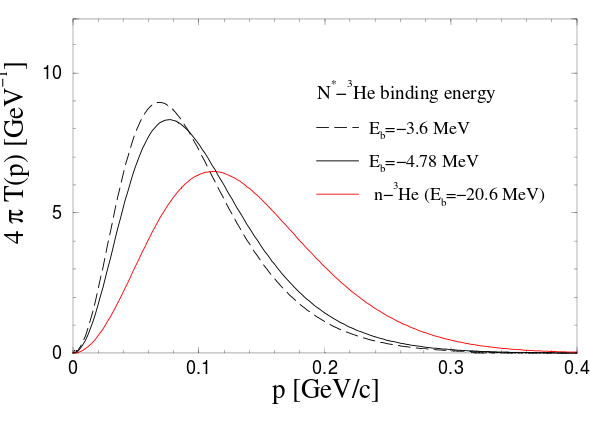}
\caption{Momentum distribution of $N^{*}$ (black solid and dashed) and neutron (red solid) inside $^{4}\hspace{-0.03cm}\mbox{He}$ nucleus calculated for $N^{*}$-$^{3}\hspace{-0.03cm}\mbox{He}$ potential for binding energy -3.6~MeV and -4.78~MeV~\cite{Kelkar_2015_new,Kelkar_2016_new} and $n$-$^{3}\hspace{-0.03cm}\mbox{He}$ potential with 20.6 MeV binding energy, respectively.~\label{Nstar_distr}}  
\end{figure}

Another contribution is connected with the assumption that the 
$N^{*}$ resonance has a momentum distribution identical to the distribution of nucleons inside Helium, which was used in the simulations of the bound state creation and decay. The application of momentum distributions based on two different potential models AV18-TM or CDB2000-UIX~\cite{Nogga2} causes only slight changes in the acceptance for simultaneous registration of all particles in the WASA detector (about 1\%).
Though the acceptance within the two different models of the nucleon distribution 
changes only slightly, it is important to test the validity of assuming a nucleon 
momentum distribution in place of that of an $N^{*}$ inside the nucleus. 
With this in mind, 
the first attempt for the evaluation of the $N^{*}$-nucleus 
potentials was performed in Ref.~\cite{Kelkar_2015_new}. The elementary 
$N N^{*} \to N N^{*}$ interaction was constructed within a $\pi$ plus $\eta$ 
meson exchange model and the $N^{*}$-nucleus potential was then obtained by 
folding the elementary $N N^{*}$ interaction with a nuclear density. A couple of 
possible bound states of the  $N^{*}$-$^{3}\hspace{-0.03cm}\mbox{He}$ system, depending 
on the choice of the $\pi N N^{*}$ and $\eta N N^{*}$ coupling constants were predicted.
This work was further extended to evaluate the bound state wave function and the 
momentum distribution of the $N^{*}$ in nuclei ~\cite{Kelkar_2016_new}. 
The $N^{*}$- $^{3}\hspace{-0.03cm}\mbox{He}$ momentum distribution for a
binding energy of -4.78~MeV and -3.6~MeV are shown in Fig.~\ref{Nstar_distr}.

These distributions are peaked at lower momentum values with respect to the distribution of 
a neutron in $^{4}\hspace{-0.03cm}\mbox{He}$ (red line) and hence leads to a 
lower acceptance because more $^{3}\hspace{-0.03cm}\mbox{He}$ nuclei will fly inside a beam pipe and will not be detected in the Forward Detector. Using Monte Carlo Simulations we estimated that the acceptance calculated assuming the $^{3}\hspace{-0.03cm}\mbox{He}$ momentum distribution indicated by the dashed line is by about 41\% smaller than the acceptance calculated assuming the distribution presented by the red solid line.

Thus when assuming in the analysis the Fermi momentum of $N^{*}$ in the $N^{*}$-$^{3}\hspace{-0.03cm}\mbox{He}$ system~\cite{Kelkar_2016_new} the estimated upper limits vary from 4.2 to 5.9~nb for the $dd\rightarrow(^{4}\hspace{-0.03cm}\mbox{He}$-$\eta)_{bound}\rightarrow$ $^{3}\hspace{-0.03cm}\mbox{He} n \pi{}^{0}$ process and from 8.5 to 11.9~nb for the $dd\rightarrow(^{4}\hspace{-0.03cm}\mbox{He}$-$\eta)_{bound}\rightarrow$ $^{3}\hspace{-0.03cm}\mbox{He} p \pi{}^{-}$, respectively.

The total systematic error was determined by adding in quadrature all contributions described above and it varies from 42\% to 46\% for $dd\rightarrow(^{4}\hspace{-0.03cm}\mbox{He}$-$\eta)_{bound}\rightarrow$ $^{3}\hspace{-0.03cm}\mbox{He} n \pi{}^{0}$ and $dd\rightarrow(^{4}\hspace{-0.03cm}\mbox{He}$-$\eta)_{bound}\rightarrow$ $^{3}\hspace{-0.03cm}\mbox{He} p \pi{}^{-}$ reaction.~The systematic uncertainties are presented by the green area in Fig.~\ref{Result_sigma_upp_both}.


\section{Summary and Perspectives}

The experiment dedicated to search for $\eta$-mesic $^{4}\hspace{-0.03cm}\mbox{He}$ in $dd\rightarrow$ $^{3}\hspace{-0.03cm}\mbox{He} n \pi{}^{0}$ and $dd\rightarrow$ $^{3}\hspace{-0.03cm}\mbox{He} p \pi{}^{-}$ reactions was performed with the WASA-at-COSY detection setup using the unique ramped beam technique.~This method allowed to change the deuteron beam momentum slowly and continuously around the $\eta$ production threshold during each of the acceleration cycles.~The acceleration covered the beam momentum range from 2.127~GeV/c to 2.422~GeV/c corresponding to the excess energy range of \mbox{$Q\in$~(-70,30)~MeV}.

The excitation functions determined for the $dd\rightarrow(^{4}\hspace{-0.03cm}\mbox{He}$-$\eta)_{bound}\rightarrow$ $^{3}\hspace{-0.03cm}\mbox{He} p \pi{}^{-}$ and the $dd\rightarrow(^{4}\hspace{-0.03cm}\mbox{He}$-$\eta)_{bound}\rightarrow$ $^{3}\hspace{-0.03cm}\mbox{He} n \pi{}^{0}$ processes do not reveal any structure which could be interpreted as a signature of a narrow bound state having a width larger than 5~MeV and smaller than 50~MeV. Upper limits of the total cross sections for the $\eta$-mesic bound state formation and decay were estimated. A simultaneous fit to excitation functions for both processes results in the value of the upper limit in the range from 2.5 to 3.5 nb for the $dd\rightarrow(^{4}\hspace{-0.03cm}\mbox{He}$-$\eta)_{bound}\rightarrow$ $^{3}\hspace{-0.03cm}\mbox{He} n \pi{}^{0}$ process and from 5 to 7 nb for the $dd\rightarrow(^{4}\hspace{-0.03cm}\mbox{He}$-$\eta)_{bound}\rightarrow$ $^{3}\hspace{-0.03cm}\mbox{He} p \pi{}^{-}$ reaction, when assuming that the momentum distribution of $N^{*}$ in the $N^{*}$-$^{3}\hspace{-0.03cm}\mbox{He}$ system is the same as momentum distribution of nucleons in the $^{4}\hspace{-0.03cm}\mbox{He}$ nucleus. However, these upper limits increase by the factor of 1.7 when assuming in the analysis that the $N^{*}$ momentum distribution is given as given by the results of the recently proposed model~\cite{Kelkar_2016_new}.

The excitation function for the $dd\rightarrow(^{4}\hspace{-0.03cm}\mbox{He}$-$\eta)_{bound}\rightarrow$ $^{3}\hspace{-0.03cm}\mbox{He} n \pi{}^{0}$ process was for the first time obtained experimentally.~The result obtained for the $dd\rightarrow(^{4}\hspace{-0.03cm}\mbox{He}$-$\eta)_{bound}\rightarrow$ $^{3}\hspace{-0.03cm}\mbox{He} p \pi{}^{-}$ reaction is about four times lower in comparison with the result obtained in a previous experiment~\cite{Adlarson_2013} and comparable with theoretical predictions resulting in $\sigma_{tot}$=4.5~nb~\cite{WycechKrzemien}.~To sum up, we may conclude  that the data collected with the WASA-at-COSY detector in 2010 do not reveal a signal for a narrow $^{4}\hspace{-0.03cm}\mbox{He}$-$\eta$ mesic nucleus.

\section*{Acknowledgements} 
The work is based on Doctoral Thesis by Magdalena Skurzok.~We acknowledge support by the Foundation for Polish Science - MPD program, co-financed by the European Union within the European Regional Development Fund, by the Polish National Science Center through grants No.~UMO-2014/15/N/ST2/03179, ~DEC-2013/11/N/ST2/04152, 2011/01/B/ST2/00431, 2011/03/B/ST2/01847 and by the FFE grants of the Forschungszentrum J\"ulich.

\end{document}